\documentclass[aps,twocolumn,groupedaddress]{revtex4}

\usepackage{graphicx}

\newcommand{ \be }{\begin{equation}}
\newcommand{ \ee }{\end{equation}}
\newcommand{ \bea }{\begin{eqnarray}}
\newcommand{ \eea }{\end{eqnarray}}
\newcommand{ \la }{\langle}
\newcommand{ \ra }{\rangle}
\newcommand{ \1 }{$v_1$}
\newcommand{ \2 }{$v_2$}

\begin{document}
\voffset=0.5 in

\preprint{LBNL-49056}

\title{Anisotropic Flow at the SPS and RHIC}


\author{Arthur M. Poskanzer}
\email[]{AMPoskanzer@LBL.gov}
\affiliation{Lawrence Berkeley National Laboratory}


\date{\today}

\begin{abstract}
The results on directed and elliptic flow for $Pb + Pb$ at the full
energy of the SPS ($158$~GeV/A) and from the first year of $Au + Au$
at RHIC ($\sqrt{s_{_{NN}}}=130$~GeV) are reviewed. The different
experiments agree well and a consistent picture has emerged indicating
early time thermalization at RHIC.
\end{abstract}


\maketitle


Elliptic flow from non-central collisions has become a prime hadronic
signature for pressure at early times in relativistic heavy ion
collisions~\cite{Sorge,Olli92}. Parton Cascade Model
calculations~\cite{PCM} show that the elliptic flow builds up in the
first few $fm/c$ and then remains constant. In this model the amount
of elliptic flow produced is proportional to the assumed parton-parton
scattering cross section. Thus it is thought that rescattering
converts the initial space anisotropy of the overlap region of
non-central collision to the momentum anisotropy of elliptic flow. As
the initial lens-shaped overlap region expands it becomes more
spherical, quenching the driving force that produces the elliptic
flow. Thus elliptic flow is sensitive to the number of interactions
and is considered to be a measure of the degree of thermalization at
early time.

\section{Introduction}
Directed and elliptic flow are defined by the \1 and \2 coefficients
in the Fourier expansion of the azimuthal distribution of particles
relative to the reaction plane:
\be
  1 + 2 v_1 \cos(\phi-\Psi_r) + 2 v_2 \cos(2(\phi-\Psi_r)) ,
\label{Fourier}
\ee
where $\phi$ is the azimuthal angle of a particle and $\Psi_r$ is the
azimuthal angle of the reaction plane.  The coefficients are usually
evaluated differentially as a function of $y$, $p_t$, and centrality.

Three methods of analysis will be discussed. In the conventional
method~\cite{meth} of correlating particles with an event plane the
flow coefficients are given by:
\be
  v_{n}^{obs} = \la \cos(n(\phi_i-\Psi_n)) \ra ,
\label{conventional}
\ee
where $\Psi_n$ is the observed event plane of order $n$.  Since the
observed event plane is not the true reaction plane, the observed
coefficients have to be corrected by dividing by the resolution of the
event plane, which is estimated by measuring the correlation of the
event planes of subevents.

The flow coefficients can also be obtained by the pair-wise
correlation~\cite{Keane} of all the particles without refering to an
event plane. This two-particle correlation method produces the squares
of the coefficients, so that one has to take the square root of the
correlation effect:
\be
  v_{n}^{2} = \la \cos(n(\phi_i-\phi_j)) \ra _{i \ne j} .
\label{pairwise}
\ee

Recently, a multiparticle correlation method has been
proposed~\cite{new-new} using cumulants. For four particles, one
calculates the four-particle correlation minus twice the square of the
two-particle correlations:
\bea
  v_{n}^{4} = & - \la \cos(n(\phi_1+\phi_2-\phi_3-\phi_4)) \ra \nonumber 
\\
  & + \la \cos(n(\phi_1-\phi_3)) \ra \ \la \cos(n(\phi_2-\phi_4)) \ra
\nonumber \\
  & + \la \cos(n(\phi_1-\phi_4)) \ra \ \la \cos(n(\phi_2-\phi_3)) \ra .
\label{cumulant}
\eea
Here one must take the fourth root of the result, and naturally the
statistical errors are larger than for the two-particle
analysis. However, this method has the advantage of eliminating
two-particle non-flow effects which are not correlated with the
reaction plane, such as HBT and resonance decay.


\section{SPS}

\subsection{NA49}

\begin{figure}[ht]
\includegraphics[width=.49\textwidth]{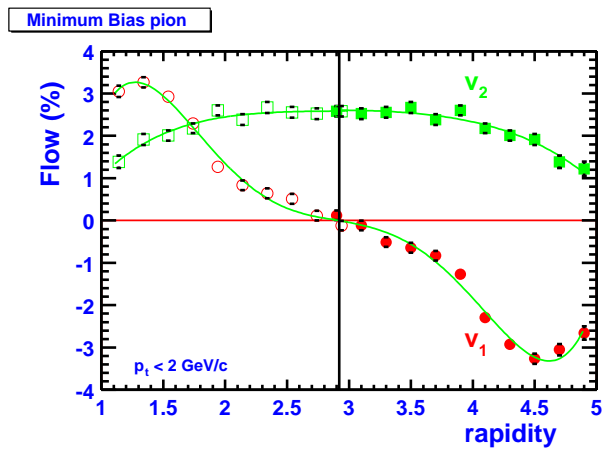}
\caption{\label{NA49mbY}NA49 directed and elliptic flow for pions from a
minimum bias trigger as a function of rapidity. The open points are
reflected. The data are still preliminary.}
\end{figure}

\begin{figure}[ht]
\includegraphics[width=.49\textwidth]{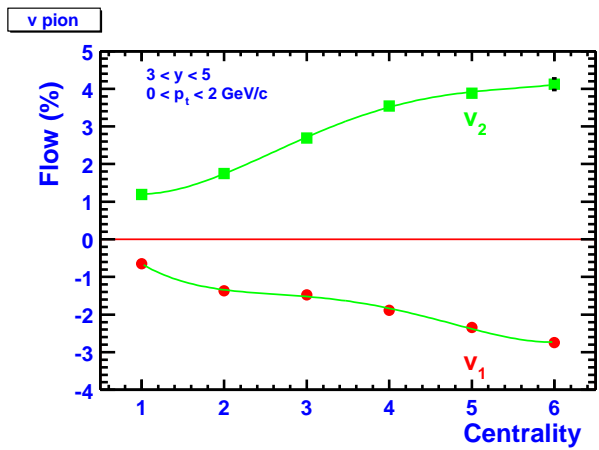}
\caption{\label{NA49vCen}NA49 directed and elliptic flow for pions as a 
function of centrality bin. The centrality increases to the left. The
data are still preliminary.}
\end{figure}

From NA49, initial flow results from part of the data set were
published~\cite{NA49PRL,NA49QM99} a few years ago. The full set of
data has now been analyzed~\cite{NA49Pri}. The \1 and \2 values for
pions as a function of rapidity are shown in Fig.~\ref{NA49mbY}. The
values have been integrated over $p_t$ by taking mean values weighted
with the cross section. The \1 values have been corrected for
conservation of momentum~\cite{OlliMomen}. Before this correction, \1
crossed mid-rapidity at +0.6\%, while after the correction, with no
adjustable parameters, it can be seen that the curve now crosses
mid-rapidity at zero. Other non-flow effects~\cite{OlliHBT,OlliSPS}
have been verified to be small using the following tests. Short-range
correlations, like HBT, were determined to be not large by using
striped sub-events where only physically separated particles were
correlated. Coulomb effects were determined to be not large by looking
at $\pi^+$ and $\pi^-$ particles separately. Neutral resonances, like
$\rho^0$, were shown to be not important by correlating $\pi^-$ with
an event plane determined only by other $\pi^-$ particles. The pion
flow values integrated over both $p_t$ and $y$ are show in
Fig.~\ref{NA49vCen} as a function of centrality bin.

\subsection{WA98}

\begin{figure}[ht]
\includegraphics[width=.49\textwidth]{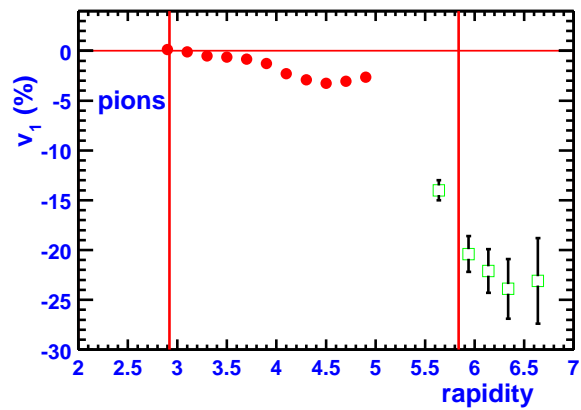}
\caption{\label{WA98}NA49 (circles) and WA98 (squares) pion data are compared. 
Plotted is the directed flow versus rapidity. The vertical lines are
at mid-rapidity and beam rapidity. The data are preliminary.}
\end{figure}

WA98 has measured~\cite{WA98QM99} the directed flow of pions near
target rapidity in the Plastic Ball. These data are plotted in
Fig.~\ref{WA98} reflected into the forward hemisphere, together with
the NA49 data. It can be seen that at projectile rapidities the
directed flow is huge. WA98 also has shown that for protons \1 is just
as large, but in the positive direction. This indicates that looking
at projectile rapidity fragments might be a good way to determine the
reaction plane. However, even though \1 is large, the resolution of
the event plane also depends on the square root of the multiplicity,
which in this rapidity region is low.

\subsection{NA45}

\begin{figure}[ht]
\includegraphics[width=.49\textwidth]{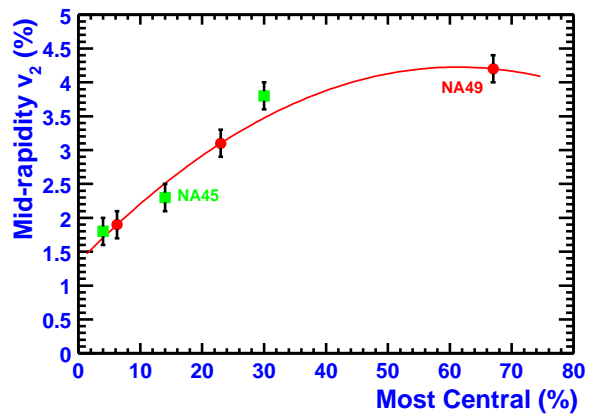}
\caption{\label{NA45}NA49 (circles) and NA45 (squares) charged
particle data are compared. Plotted is the elliptic flow at
mid-rapidity versus the centrality. The centrality increases to the
left because the abscissa is the cumulative most central percentage of
all events that are in the bin. The data are preliminary.}
\end{figure}

Using silicon drift chambers NA45 has measured~\cite{NA45QM01} the
flow coefficients for charged particles. In Fig.~\ref{NA45} their
elliptic flow values at mid-rapidity are plotted versus centrality,
together with the same quantity for pions from NA49. The agreement is
good. NA45 also has a RICH detector which identifies high $p_t$
pions. When correlated with an event plane determined from their TPC,
the $p_t$ dependence falls nicely on a line extrapolated from the
lower $p_t$ results of NA49. However, when they correlate pairs of
high $p_t$ pions from the RICH using Eq.~\ref{pairwise}, their results
are somewhat higher~\cite{NA45INT}.

\subsection{NA50}

NA50 has analyzed the neutral transverse energy from their calorimeter
in the rapidity region from 1.1 to 2.3. NA50 has shown elliptic flow
as a function of centrality~\cite{NA50QM99}, but it is hard to compare
their results with the other experiments because of the different method of
analysis.

\section{RHIC}

\subsection{STAR}

STAR results for elliptic flow for charged
particles~\cite{STARcharged} and identified particles~\cite{STARPID}
have been published. The flow is larger than at the SPS and comes
close to the hydrodynamic predictions~\cite{Pasi} for the mid-central
and central data. The identified particle flow shows a variation with
mass as predicted by hydrodynamics. However, the high $p_t$
behavior~\cite{STARQM01} exhibits a saturation effect well below where
hydro would saturate. It has been suggested that at high $p_t$
elliptic flow could be due to parton energy loss (jet
quenching)~\cite{Gyulassy}.

\begin{figure}[ht]
\includegraphics[width=.49\textwidth]{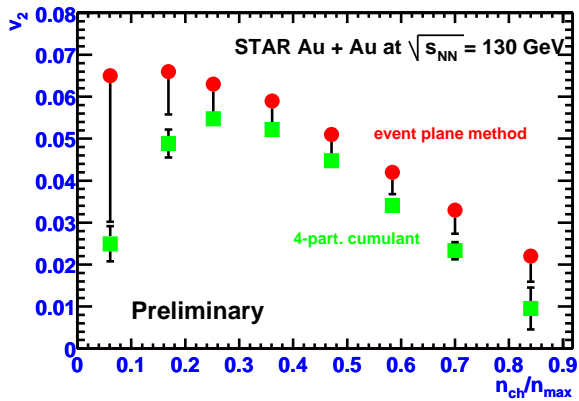}
\caption{\label{STAR}STAR elliptic flow for charged particles versus 
centrality. The results using the conventional event plane method
(circles) and the four-particle cumulant method (squares) are
shown. The later are preliminary.}
\end{figure}

The doubly integrated elliptic flow values as a function of centrality
are shown in Fig.~\ref{STAR}, using the conventional event plane
method (Eq.~\ref{conventional}). The asymmetric error bars indicate
the systematic uncertainty due to possible non-flow effects.  A
four-particle correlation analysis~\cite{STARAihong} using
Eq.~\ref{cumulant} is also show in the figure. Indeed, when
two-particle non-flow effects are eliminated, the points do come down
to about the limits of the systematic errors. It should be noted that
part of the decrease could be due to fluctuation effects~\cite{fluct}
which also would go in the same direction. However, the agreement with
hydro for mid-central and central collisions is not affected.

\subsection{PHENIX}

\begin{figure}[ht]
\includegraphics[width=.49\textwidth]{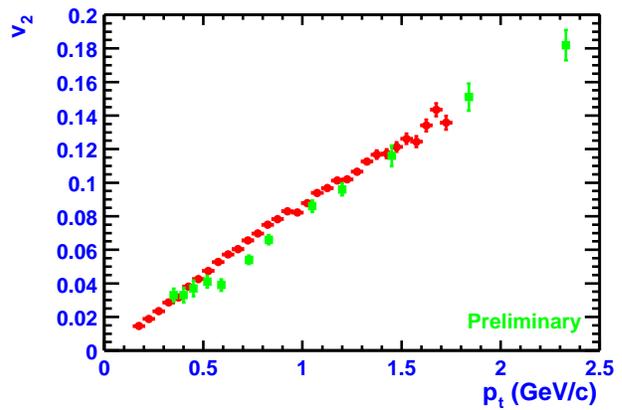}
\caption{\label{PHENIX}STAR (circles) and PHENIX (squares) charged 
particle data for a minimum bias trigger are compared. Plotted is the
elliptic flow versus the transverse momentum. The PHENIX data are
preliminary.}
\end{figure}

The PHENIX results~\cite{PHENIX} for charged particle elliptic flow as
a function $p_t$ for a minimum bias trigger is shown together with the
STAR results in Fig.~\ref{PHENIX}. PHENIX uses the pair-wise method of
Eq.~\ref{pairwise}. The agreement is excellent. However, at the
moment, the doubly-integrated elliptic flow values as a function of
centrality are about 50\% higher for PHENIX than for STAR. Only about
half of this discrepancy can be accounted for by the higher $p_t$
threshold of PHENIX.

\subsection{PHOBOS}

\begin{figure}[ht]
\includegraphics[width=.49\textwidth]{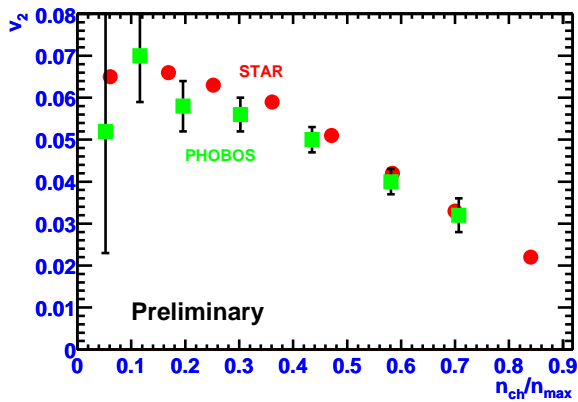}
\caption{\label{PHOBOScen}STAR (circles) and PHOBOS (squares) charged 
particle data are compared. Plotted is the elliptic flow versus
centrality. The PHOBOS data are preliminary and the error bars are
only statistical.}
\end{figure}

\begin{figure}[ht]
\includegraphics[width=.40\textwidth]{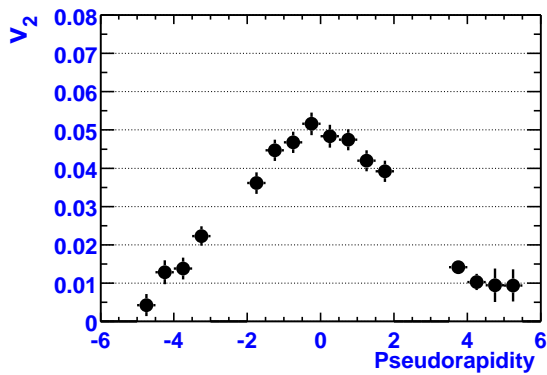}
\caption{\label{PHOBOSeta}PHOBOS charged particle elliptic flow for a
minimum bias trigger plotted versus pseudorapidity. The data are
preliminary and the error bars are only statistical.}
\end{figure}

The PHOBOS elliptic flow values~\cite{PHOBOS} as a function of
centrality are shown in Fig.~\ref{PHOBOScen}, together with the STAR
values. The agreement is excellent. Fig.~\ref{PHOBOSeta} shows PHOBOS
elliptic flow for a minimum bias trigger measured over a wide range of
pseudorapidity. Where there are STAR measurements, in a narrow range
near mid-rapidity, the agreement is good with the flat STAR
distribution. A 3D hydro calculation~\cite{Hirano} gets rather flat
distributions over a wide range of pseudorapidity for all reasonable
initial conditions. Thus the fast fall off away from midrapidity in
Fig.~\ref{PHOBOSeta} is not understood at present in a model with
thermalization.

\section{Conclusions}

Good agreement between the various experiments has been
achieved. There might be some indication that the pair-wise method for
high $p_t$ particles indicates non-flow effects. The four-particle
method at RHIC indicates that there are non-flow effects for
peripheral collisions. The elliptic flow increases from the SPS to
RHIC, and at RHIC is close to the hydro limit. This is taken to be a
signature for early time thermalization at RHIC.


\begin{acknowledgments}
 I wish to thank Alexander Wetzler, Thomas Peitzmann, Kirill
 Filimonov, Roy Lacey, Steve Manly, Inkyu Park, and Peter Steinberg
 for supplying some of the data and graphs. Comments by Hans-Georg
 Ritter, Nu Xu, Sergei Voloshin, Jean-Yves Ollitrault, and Peter
 Seyboth are also appreciated. This work was supported by the Division
 of Nuclear Physics of the Office of Science of the U.S.Department of
 Energy.
\end{acknowledgments}


\end{document}